# Thermal conductivity prediction of nanoparticle packed beds by using modified Johnson-Kendall-Roberts model

Zizhen Lin and Congliang Huang

School of Electrical and Power Engineering, China University of Mining and Technology, Xuzhou 221116, China.

**Abstract** Nanoparticle packed beds (NPBs) have demonstrated the potential for thermal insulation, and further reducing thermal conductivity ($k$) requires a theoretical understanding of the thermal conduction in them. Till now, the theoretical models under the framework of effective medium approach (EMA) have been widely developed for the thermal conductivity ($k$) prediction of NPB. In these models, corresponding architecture parameters are usually evaluated by the classical Johnson-Kendall-Roberts (JKR) model. Unfortunately, the size effect is usually ignored in JKR model, resulting in the inferior ability to accurately predict the geometrical information of NPB. In this work, the modified JKR model including the size effect of Young's modulus is integrated in the EMA model for $k$ prediction, and experimental results in [Int. J. Heat Mass Tran., 2019, 129, 28-36] was further explained. As a result, the developed model illustrates the advantages on the prediction of solid phase thermal conductivity ($k_s$), especially for the NPB with a low porosity. This work provides a modified JKR model to improve the accuracy of existing EMA model in the $k$ prediction of NPB.

Despite the silica-based nanoporous materials have received wide concern for thermal insulation because of their low thermal conductivity ($k$) [1,2], rare silica thermal insulation materials show the $k$ lower than the Einstein limit ($k_E$) [3], which describes the lower limit of $k$ of the silica bulk with amorphous configuration. In addition, the brittle nature also restricts the application of silica nanoporous materials in several areas, such as space applications. The nanoparticle packed bed (NPB) provides an alternative choice as the next generation thermal insulation materials because of the low thermal conductivity, even lower than the free air (0.026 W m$^{-1}$ K$^{-1}$) and high strength which could be elucidated by the effect of the inter-nanoparticle contact interfaces. The high-density interfaces not only introduce an effective scattering mechanism for mid-and long-wavelength phonons for low thermal conductivity [4], but also provides the strong Van der Waals to guarantee the strong strength. To deepen the insight into the interfacial effect on the $k$, some models have already been proposed. The classical acoustic mismatch model (AMM) model and the diffuse mismatch model (DMM) are originally proposed under the assumption of the specular or diffusive phonon transport at interface. Further taking the size effect of interface into account, Prasher successfully models $k$ under the framework of EMA with the ballistic and diffusive transport behavior of interfacial phonons considered [5,6]. Moreover, the influence of nanoparticle deformation also contains in the $k$ model by coupling the classical Johnson-Kendall-Roberts (JKR) model into existing EMA model. Whereas the JKR shows the limited ability in capturing the architecture parameters of NPB, resulting in the inferior effect of existing EMA model in $k$ prediction, especially for the NPB with a low porosity ($\varphi$) [7]. In this work, the modified JKR model including the size effect of Young's modulus is integrated in the EMA model for $k$ prediction, and experimental results in Ref. [7] was further explained. As a result, the optimized theoretical model demonstrates the advantages on the prediction of the solid-phase thermal conductivity ($k_s$) than the traditional EMA model given in Ref. [7].

For simplicity, all nanoparticles within NPB are assumed to be uniformly distributed by a form of a simple cubic lattice. The radius ($a$) of the contact surface between nanoparticles can be described by the improved Johnson-Kendall-Roberts theory, [8,9]

$$[(2-\varphi)a]^3 = \frac{3\pi KrF}{4}, \quad (1)$$

where $F$ is the loaded force in the preparation of a nanoparticle packed bed. $K$ is the elastic constant of silica nanoparticles, which can be expressed as $K = \frac{1-v^2}{\pi NE}$, where $v$, $E$ and $N$ are the Poisson ratio, Young's modulus and the correction factor for the size effect of Young's modulus. The factor $N$ is about 1 for 50 nm and 0.45 for 10 nm silica nanoparticles. For NPB, the loaded pressure is a function of $\varphi$, can be expressed as $\frac{P}{P_0} = f(\varphi)$, where $P_0$ is the maximum pressure used in the bed preparation, and the $f(\varphi)$ function can be obtained by experimental methods. Here, a simple equation is applied to fit the experimental results,

$$\frac{P}{P_0} = f(\varphi) \propto \sqrt[x]{(1-\varphi)}, \tag{2}$$

where $x$ are the fitting parameter as exhibited in the inset of Fig. 1.

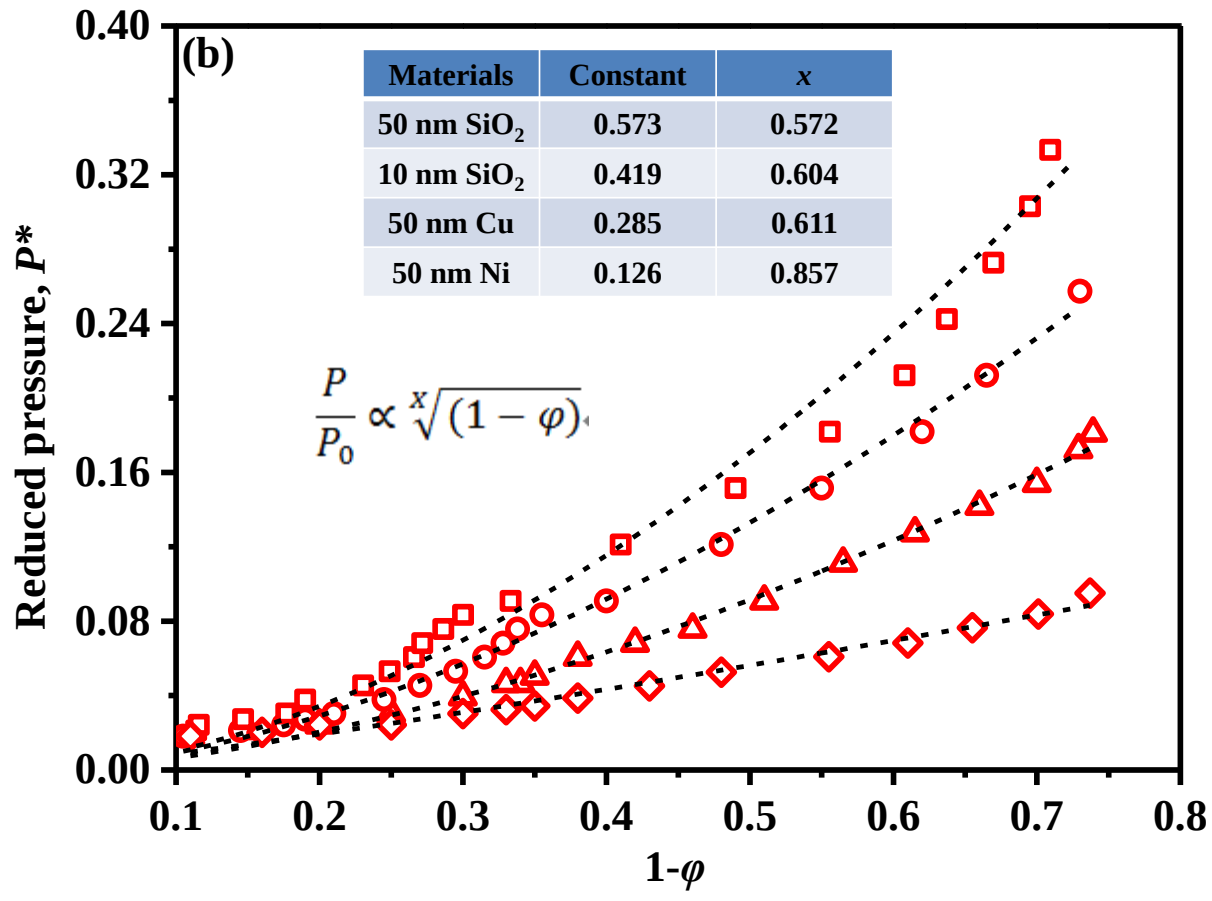

Fig. 1 Reduced pressure ($P/P_0$) as a function of ($1$-$\varphi$).

The thermal resistances ($R$) in NPB contains two parts: a) the resistance caused by the mesoscopic contact area defect ($R_a$). b) the resistance caused by the interfacial phonon scatterings ($R_c$). The phonon transport at nanoparticle interface shows the elastic scattering behavior. Thereby, $R_a$ can be contributed by two different mechanism, the one is the thermal resistance aroused by the inadequate nanoparticle contact ($R_1$), and another one is the interfacial resistances caused by the lattice defect between nanoparticles ($R_2$). Thus, $R_a$ can be expressed as [10],

$$R_a = R_1 + R_2 = \frac{\psi}{k_0 \varepsilon A_\varphi} + \left[A_\varphi \frac{\pi^2}{60} \frac{k_B^4}{\hbar^3} \upsilon^{-2} T^3\right]^{-1}, \tag{3}$$

where $k_0$ is the lattice thermal conductivity of silica nanoparticle, the $k_0$ of 50 nm and 10 nm silica nanoparticle are about 0.51 and 0.4 W m$^{-1}$ K$^{-1}$ respectively, which can be predicted by the Callaway model [11]. $\psi$ is the root mean square roughness, herein, the $\psi = 1$ nm is used for silica nanoparticle. $\varepsilon$ describes the compactness degree of nanoparticles within NPBs, $\varepsilon$=0.5 for spherical nanoparticle. $A_\varphi$ is the contact surface area, ħ is the reduced Planck constant.

Considering that the phonon mean free path ($l$) of silica nanoparticle (about 1-6 nm) is on the same order of magnitude of the character size ($a$) of contact area between nanoparticles, $R_c$ should contain the thermal resistance caused by both the interfacial phonon ballistic scattering ($R_{cb}$) and the interfacial phonon diffusive scattering ($R_{cd}$). In this work, the $R_c$ is predicted based on a classical model given by Prasher [12], $R_c = R_{cd} + R_{cb}$. The $R_{cd}$ is given by the Maxwell constriction resistance model derived from solving the heat conduction equation based on Fourier's law,

$$R_{cd} = \frac{1}{2k_0 a}. \tag{4}$$

To convenient the description of ballistic phonons across the interfacial region, the heat-driven phonon flow is deemed as a gas flow, therefore the $R_{cb}$ can be calculated by considering the flow rate of gas molecules through an orifice in the free molecular flow regime, as

$$R_{cb} = \frac{4l}{3A_\varphi k_0}, \tag{5}$$

The total thermal resistance of a particle-packed bed can be written as

$$R = \left(R_a^{-1} + R_c^{-1} + R_g^{-1}\right)^{-1} + R_{bulk}, \tag{6}$$

where $R_g$ is the thermal resistance at solid-gas interface, expressed as $R_g = \frac{2R_{s-g}}{\pi r^2}$, $R_{s-g}$=4.9×10$^{-6}$ K·m$^2$·W$^{-1}$ [13,14]. To further simplify the calculation process, a more general $k_s$ model is used [15].

$$k_s = \frac{2r}{\pi D^2} \times \frac{1}{R}, \tag{7}$$

The theoretical $k_s$ matches experimental data well as shown in Fig. 2 [7], indicating the rationality of the developed model in this work. More importantly, this model also illustrates the advantages on the $k_s$ prediction than the EMA model with the classical JKR model employed in the prediction of inter-nanoparticle contact surface area. Therefore, it is rationally deduced that the size effect of structure strength should be considered in the NPB system, and maybe the critical factor to determine the ability of theoretical model in $k_s$ prediction of NPB, especially for the one with a low porosity. In this regard, the modified JKR model would be meaningful in improving the accuracy of the $k$ prediction by the existing EMA model.

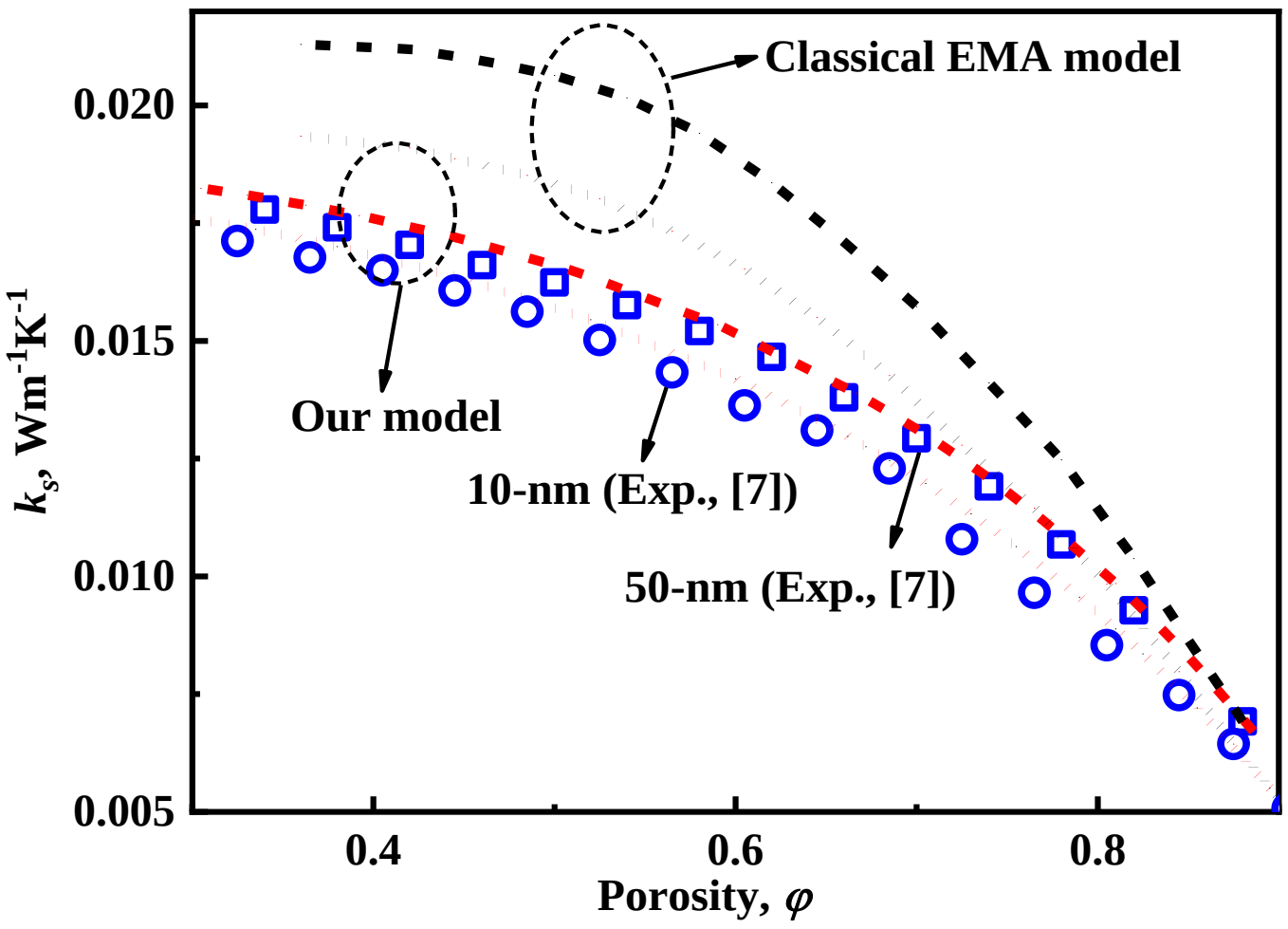

Fig. 2 Theoretical and experimental $k_s$ comparison of the 10-nm and 50 nm NPBs.